\documentclass[a4paper,11pt]{article}
\pdfoutput=1 % if your are submitting a pdflatex (i.e. if you have
\usepackage{jinstpub} % for details on the use of the package, please
\usepackage{amsmath} 
\usepackage{xfrac}
\usepackage[capitalize]{cleveref}
\usepackage{xcolor} 
\usepackage{placeins}
\usepackage{latexsym}

\renewcommand{\d}{\text{d}}
\def\*#1{\mathbf{#1}} 
\def\~#1{\mathcal{#1}}
\def\|#1{\mathbb{#1}}
\def\.#1{\dot{#1}}
\def\:#1{\ddot{#1}}
\newcommand{\tran}{^{\mkern-1.5mu\mathsf{T}}}

\begin{document}
\title{Constant-Tune Cyclotrons}
\author{Thomas Planche}
\affiliation{TRIUMF}
\abstract{In this paper we demonstrate that cyclotrons can be made to have precisely constant betatron tunes over wide energy ranges. In particular, we show that the horizontal tune can be made constant and does not have to follow the Lorentz factor $\gamma$, while still perfectly satisfying the isochronous condition. To make this demonstration we developed a technique based on the calculation of the betatron tunes entirely from the geometry of realistic non-hard-edge closed orbits. We present two particular cyclotron designs, one compact cyclotron and one ring cyclotron. The compact cyclotron design is backed up by a 3-dimensional finite element magnet calculation, that we also present here.}
\maketitle

\section{Introduction}

The essential property of a cyclotron is the timing of its orbits: they are isochronous. As the charged particles speed up, they circulate taking longer orbits.
The static magnetic field that keeps them going around is tuned precisely so that the two effects cancel. Therefore, the orbit period remains constant and the beam can be accelerated continuously, by a fixed-frequency rf system, without pulsing. Historically, this was first capitalized upon by Ernest Lawrence~\cite{lawrence1934method}.

The orbits must be maintained isochronous with a precision that is rather unforgiving. The integral of the relative error can never exceed the ratio between a quarter of the rf period and the total time of flight from injection: this is on the order of 10$^{-4}$ or less, for most cyclotrons. Stray away from this  strict condition, and no more beam will come out of the machine. Indeed, if a particle sees its rf phase slip off crest by more than 90 degrees, it will get decelerated all the way back to the energy at which it had been injected, see for instance~\cite[Fig.~2]{Craddock:1977xk}.

In addition to the isochronous condition, the static magnetic field must also ensure the transverse confinement of the beam.
%This realization led Bethe and Rose~\cite{bethe1937maximum} to predict that cyclotron would never exceed $\sim$12\,MeV. They would be later  by Thomas~\cite{thomas1938paths} who invented the ``azimuthally-varying field'' (AVF) cyclotron, and reconciliating isochronism with stable vertical motion in high-energy cyclotron.
The transverse betatron tunes can in some cases be roughly estimated using the hard-edge approximation and the concatenation of drift-edge-bend-edge-drift transfer matrices, see for instance Refs.~\cite{gordon1968orbit,TRI-BN-82-12,TRI-BN-82-20,craddock2008cyclotrons}. But those matrices are a good approximation only in the highly idealized dipole magnets used in beamlines and synchrotrons; they have constant radius of curvature and constant field index on a chosen reference orbit that the beam is intended to follow. Real-life cyclotrons are fundamentally different with both the field and its radial derivative varying at all points along a closed orbit.

In this paper we show how, in a cyclotron with mid-plane symmetry, the transverse tunes are entirely determined by the geometry of its closed orbits. We present a method to calculate tunes and produce isochronous field maps by starting from the geometry of realistic non-hard-edge orbits.
The main advantage of the proposed method is that it produces realistic and perfectly isochronous field distributions, and the corresponding transverse tunes, in a split second. We take advantage of this speed to explore the range of possible isochronous fields, which leads us to find solutions with simultaneously constant vertical and horizontal tunes over wide energy ranges. We chose one particular solution, optimized for the case of a 3-sector 200\,MeV compact H$_2^+$ cyclotron, and we present the corresponding 3-dimensional finite element magnet design.

In the last part of this paper, we present the the design of a 800\,MeV to 2\,GeV cyclotron, with long straight sections and tunes constant over the entire energy range.

\section{Theory: Tunes of Isochronous Orbits}

\subsection{Infinitesimal Transverse Motion}
The Hamiltonian that governs the linear transverse motion of particles in a static magnetic field with median plane symmetry is given by~\cite{CourantSnyder1958,TRI-DN-05-06}:
\begin{equation}\label{eq:ham}
    h(x,p_x,y,p_y;s)=\frac{x^2}{2}\frac{1-n}{\rho^2}+\frac{y^2}{2}\frac{n}{\rho^2}+\frac{p_x^2}{2}+\frac{p_y^2}{2}\,,
\end{equation}
where $\rho$ is the local curvature of the orbit and $n=-\frac{\rho}{B_0} \left. \frac{\partial B}{\partial x} \right|_{x=y=0}$ is the local field index. The Frenet-Serret coordinates are $x$, $y$, $s$ and the canonically conjugated momenta $p_x$, $p_y$, and $h$ have all been normalized by the reference particle momentum $P=qB_0\rho$. Note that~\cref{eq:ham} is derived from the more general Courant-Snyder Hamiltonian~\cite[Appendix B]{CourantSnyder1958} after imposing that the static magnetic field has mid-plane symmetry and satisfies Maxwell's equations in their homogenous forms. It is no different from the Hamiltonian for separated function synchrotrons; for example in quadrupoles, $\frac{-n}{\rho^2}$ is replaced by its limit as $\rho\rightarrow\infty$, namely, $\frac{1}{B_0\rho}\frac{\partial B}{\partial x}$.

The equations of motion that derive from this Hamiltonian can be written in matrix form:
\begin{equation}\label{eq:eom-midplane}
    \*X'= \*F \*X\,,
\end{equation}
where a prime $'$ denotes a total derivative w.r.t.~the independent variable $s$. The matrix $\*F$:
\begin{equation}\label{eq:F-midplane}
    \*F
    = \left(
    \begin{array}{cccccc}
            0                        & 1 & 0                       & 0 \\
            \frac{n(s)-1}{\rho(s)^2} & 0 & 0                       & 0 \\
            0                        & 0 & 0                       & 1 \\
            0                        & 0 & -\frac{n(s)}{\rho(s)^2} & 0 \\
        \end{array}
    \right)
\end{equation}
is the infinitesimal transfer matrix~\cite{sacherer1971rms}, and
\begin{equation}
    \*X=
    \left(
    \begin{array}{c}
        x   \\
        p_x \\
        y   \\
        p_y
    \end{array}
    \right)
    =
    \left(
    \begin{array}{c}
        x  \\
        x' \\
        y  \\
        y'
    \end{array}
    \right)
\end{equation}
is the particle state vector.
Transverse tunes are obtained by numerically integrating:
\begin{equation}\label{eq:eom-theta}
    \frac{\d \*X}{\d \theta} = \*X' \frac{\d s}{\d \theta} =  \*F \*X \frac{\d s}{\d \theta}
\end{equation}
over one period for two different sets of initial state vectors: $\*X = (1,0,1,0)\tran$ and $\*X = (0,1,0,1)\tran$, see for instance Ref.~\cite{TRI-DN-70-54-Add4}.
One only needs to know how to calculate the field index $n$, the radius of curvature $\rho$, and $\frac{\d s}{\d \theta}$ at any point along an orbit to track the linear motion of particles around it, and obtain transverse tunes exact to arbitrary precision. Let us now see how these can be calculated from the geometry of the closed orbits.

% This Hamiltonian is all we need to calculate transverse tunes. It is entirely determined by the curvature $\rho$ and the field index $n$, which are in general a function of the independent variable $s$. Let us now see how these can be calculated from the geometry of the closed orbits.

\subsection{Parametrization of the Closed Obits}
Instead of starting from a magnetic field distribution to compute the properties of the orbits, let's start from the geometry of the orbits represented by:
% \begin{equation}\label{eq:r-general}
%     r(a,\theta)\colon \|R^{+}\times\|R \to \|R^{+}\,,
% \end{equation}
\begin{equation}\label{eq:r-general}
    \begin{aligned}
        r\colon \|R^{+}\times\|R &\to \|R^{+}\\
        (a,\theta) &\mapsto r(a,\theta)\,,
    \end{aligned}
\end{equation}
where $r$ is the radius of the closed orbit, $a$ is the orbit's average radius, and $\theta$ is the azimuth.
The periodicity of the closed orbits imposes that:
\begin{equation}\label{eq:r-periodicity}
    r(a,\theta)=r(a,\theta+2\pi/N)\text{ with } N\in\|N^*\,,
\end{equation}
where $N$ is the lattice periodicity, i.e.~number of sectors. Note that, because of their periodicity, any ensemble of closed orbits can be written in the form of a Fourier series:
\begin{equation}\label{eq:fourier}
    r(a,\theta) = aC_0(a) + a\sum_{j=1}^\infty C_j(a)\cos(jN\theta) + S_j(a)\sin(jN\theta)\,,
\end{equation}
with the Fourier coefficients $C_j(a)$ and $S_j(a)$ functions of the average orbit radius $a$.
Although this is not strictly necessary, we impose that:
\begin{equation}
    \frac{\partial r}{\partial a}>0\,,
\end{equation}
which guarantees that the closed orbits never cross\footnote{This can be generalized, so long as the resulting magnetic field is uniquely defined where orbits cross.}.
The last assumption is that the orbits are
isochronous\footnote{The more general case of non-isochronous orbits, where $\beta(a)$ can be arbitrary, is treated here~\cite{planche2019designing}.},
which leads to the following relation for the particle's relative speed $\beta$:
\begin{equation}\label{eq:beta}
    \beta(a) = \frac{\mathcal{R}(a)}{\mathcal{R}_\infty}\,,
\end{equation}
where $\mathcal{R}_\infty$ is a constant, corresponding to the value of $\mathcal{R}$ in the limit that the particles' speed is light speed; and $\mathcal{R}$ is the orbit circumference divided by $2\pi$:
\begin{equation}
    \mathcal{R}(a) = \frac{1}{2\pi}\int_0^{2\pi} \frac{\d s}{\d\theta}\,\,\d\theta\,,
\end{equation}
and:
\begin{equation}\label{eq:dsdth}
    \frac{\d s}{\d\theta} = \sqrt{r^2+\left( \frac{\partial r}{\partial \theta} \right)^2}\,.
\end{equation}
The local curvature of the orbit is given by the standard formula for polar coordinates:
\begin{equation}\label{eq:rho}
    \frac{1}{\rho} = \frac{r^2+2\left( \frac{\partial r}{\partial \theta} \right)^2-r\frac{\partial^2 r}{\partial \theta^2}}{\left( r^2+ \left( \frac{\partial r}{\partial \theta} \right)^2 \right)^{3/2}}\,.
\end{equation}
The field index $n$ is obtained from:
\begin{equation}
    n=-\frac{\rho}{B_0} \frac{\partial B}{\partial x} = \frac{\partial \rho}{\partial x} - \frac{\rho}{\beta\gamma}\frac{\partial \beta\gamma}{\partial x}
\end{equation}
where $\beta\gamma = \frac{\beta}{\sqrt{1-\beta^2}}$ and $\beta$ is given by~\cref{eq:beta}.
\begin{figure}[!htb]
    \centering
    \includegraphics*[width=0.5\columnwidth, trim={60 0 0 0},clip]{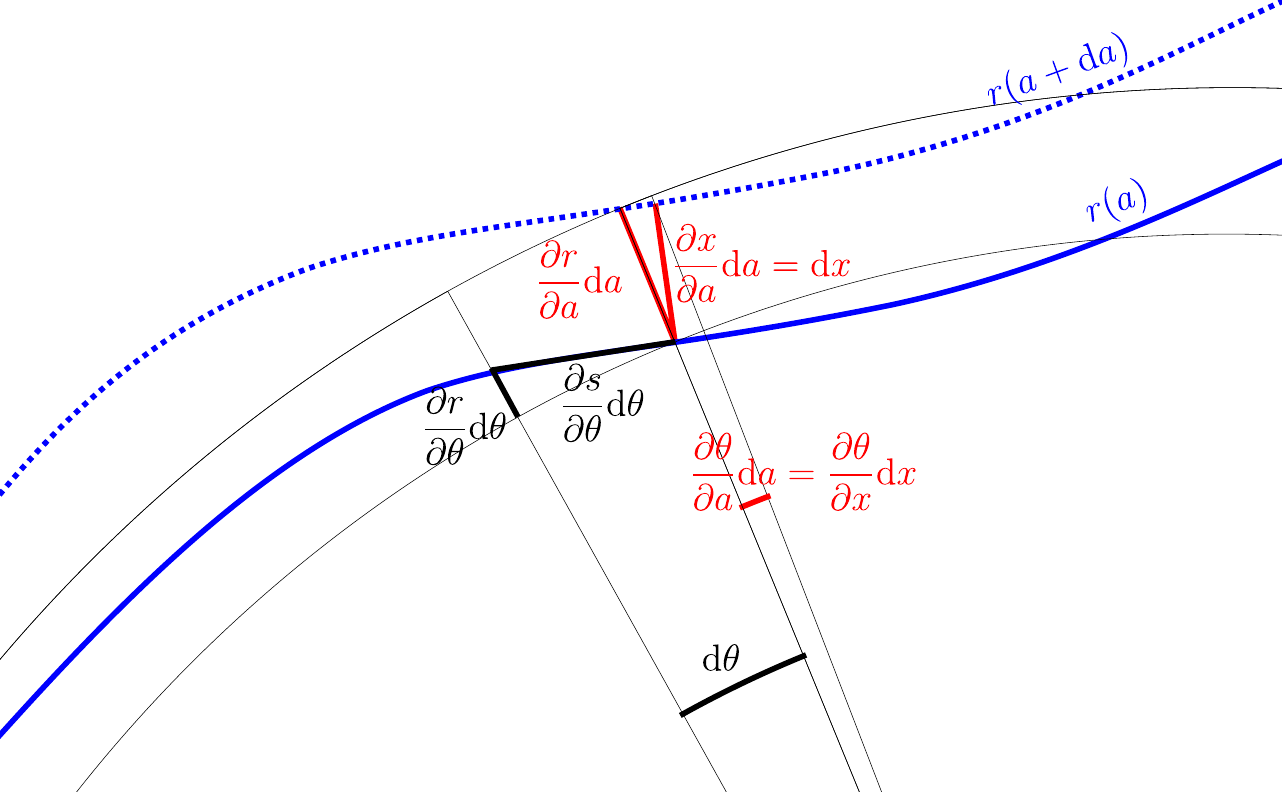}
    \centering
    \caption{Relations between infinitesimal quantities around the closed orbit (the thick blue line); d$\theta$ and d$a$ represent infinitesimal variations in $\theta$ and $a$ respectively.}
    \label{fig:traj}
\end{figure}
As shown in Ref.~\cite{planche2019designing} the chain rule and the relations between infinitesimal quantities illustrated in~\cref{fig:traj} yields:
\begin{equation}\label{eq:drhodx}
    \frac{\partial \rho }{\partial x} =
    \frac{\partial \rho }{\partial a}\frac{\partial a }{\partial x} +\frac{\partial \rho }{\partial \theta}\frac{\partial \theta }{\partial x}
    = \frac{1}{r}\left(\frac{\partial \rho }{\partial a} \frac{\frac{\d s}{\d\theta}}{\frac{\partial r}{\partial a}}
    - \frac{\partial \rho }{\partial \theta}\frac{\frac{\partial r}{\partial \theta}}{\frac{\d s}{\d\theta}}\right)\,,
\end{equation}
where $\frac{\d s}{\d\theta}$ is given by~\cref{eq:dsdth}.
If one chooses $\mathcal{R}_\infty$ to be the unit of length, the numerical integration is done entirely from the knowledge of $r(a,\theta)$ and its partial derivatives. And since the transverse tunes are unitless numbers, their values are not affected by the choice of the unit of length, i.e.\ independent of $\mathcal{R}_\infty$. The transverse tunes derive entirely from the shape of the orbits, and from absolutely nothing else.

Note that, to be able to calculate tunes in this way, the function $r(a,\theta)$ must be sufficiently smooth for the following partial derivatives: $\frac{\partial r}{\partial \theta}, \frac{\partial r}{\partial^2 \theta^2},\frac{\partial^3 r}{\partial \theta^3}, \frac{\partial r}{\partial a}, \frac{\partial^2 r}{\partial a\theta}, \frac{\partial^3 r}{\partial a\theta^2}$  to be defined.

\subsection{Corresponding Isochronous Field Map}
So far we have only needed the knowledge of the shape $r(a,\theta)$ of the closed orbits to calculate tunes. The question that comes to mind is: what magnetic field distribution produces these closed orbits? The answer will depend on the choice of the particle mass $m$ and charge $q$, and the scale of the solution will depend on the value of $\mathcal{R}_\infty$.
The magnetic field within the median plane is given by:
\begin{equation}\label{eq:Brtheta}
    B(r,\theta) =\frac{\beta(a)}{\sqrt{1-\beta^2(a)}}\frac{m}{q \rho(a,\theta)}\,,
\end{equation}
where $\rho(a,\theta)$ is given by~\cref{eq:rho}, $\beta(a)$ is given by~\cref{eq:beta}, and $a=a(r,\theta)$ is calculated from $r(a,\theta)$ using numerical root finding. The magnetic field off the median plan can be obtained from extrapolation using Maxwell's equations, see for instance Ref.~\cite{zhang2022asurement}.

In the examples presented below, the transverse tunes and orbital frequency are crosschecked using particle tracking in 2-dimensional polar field map with the reference code \texttt{CYCLOPS}~\cite{gordon1984computation}. The field maps provided to \texttt{CYCLOPS} are generated using~\cref{eq:Brtheta}.

\section{Compact Cyclotron Example}

% The conventional approach to cyclotron design is to start from the geometry of the magnet, modeled in some 3-dimensional finite element code, and to modify this geometry to iteratively produce an isochronous field distribution. This typically takes many iterations of  hours-long calculation to produce 1 field map.
% By contrast, when starting from some arbitrary function $r(a,\theta)$ one can produce an isochronous field map in a split second. One can then search the large parameter space of possible $r(a,\theta)$ functions to explore, relatively much more quickly, the range of possible isochronous distributions.

\subsection{Geometrical Design}\label{seq:GeometricalDesign}
Let's consider the following way of parametrizing the shape of the closed orbits:
\begin{equation}\label{eq:rath1h}
    r(a,\theta) =  a \big(1 +  C(a)\cos\big(N(\theta - \phi(a))\big)\big)\,,
\end{equation}
which is equivalent to using~\cref{eq:fourier} truncated to order $j=1$. One now needs to find a way to define the two functions $C(a)$ and $\phi(a)$ using a finite -- and hopefully small -- number of degrees of freedom. We do this by constraining the values of $C(a)$ and $\phi(a)$ for a few orbits, and we use a cubic spline interpolation for any intermediate value of $a$.

In the example presented in~\cref{tab:orb}, we define $r(a,\theta)$ following~\cref{eq:rath1h} constraining the values of $C(a)$ and $\phi(a)$ for only 5 different orbits. Since we are considering the design of a compact cyclotron, where the radius of the innermost orbit is small compared to the magnetic gap of the cyclotron magnet, the innermost orbit is necessarily very close to a perfect circle~\cite{planche2022intensity}. For this reason we choose $C(0.004\times\mathcal{R}_\infty)=0$. The initial value of $\phi(a)$ can be chosen arbitrarily, and so we set $\phi(0.004\times\mathcal{R}_\infty)=0$. (We resign ourselves to the fact that this first turn will have radial tune of 1 and vertical tune of zero.) Four more values of $C$ and $\phi$ remain to be chosen: the magnetic field of the entire cyclotron is parametrized with only $4\times2=8$ degrees of freedom.

\begin{figure}[!htb]
    \centering
    \includegraphics*[width=0.7\columnwidth]{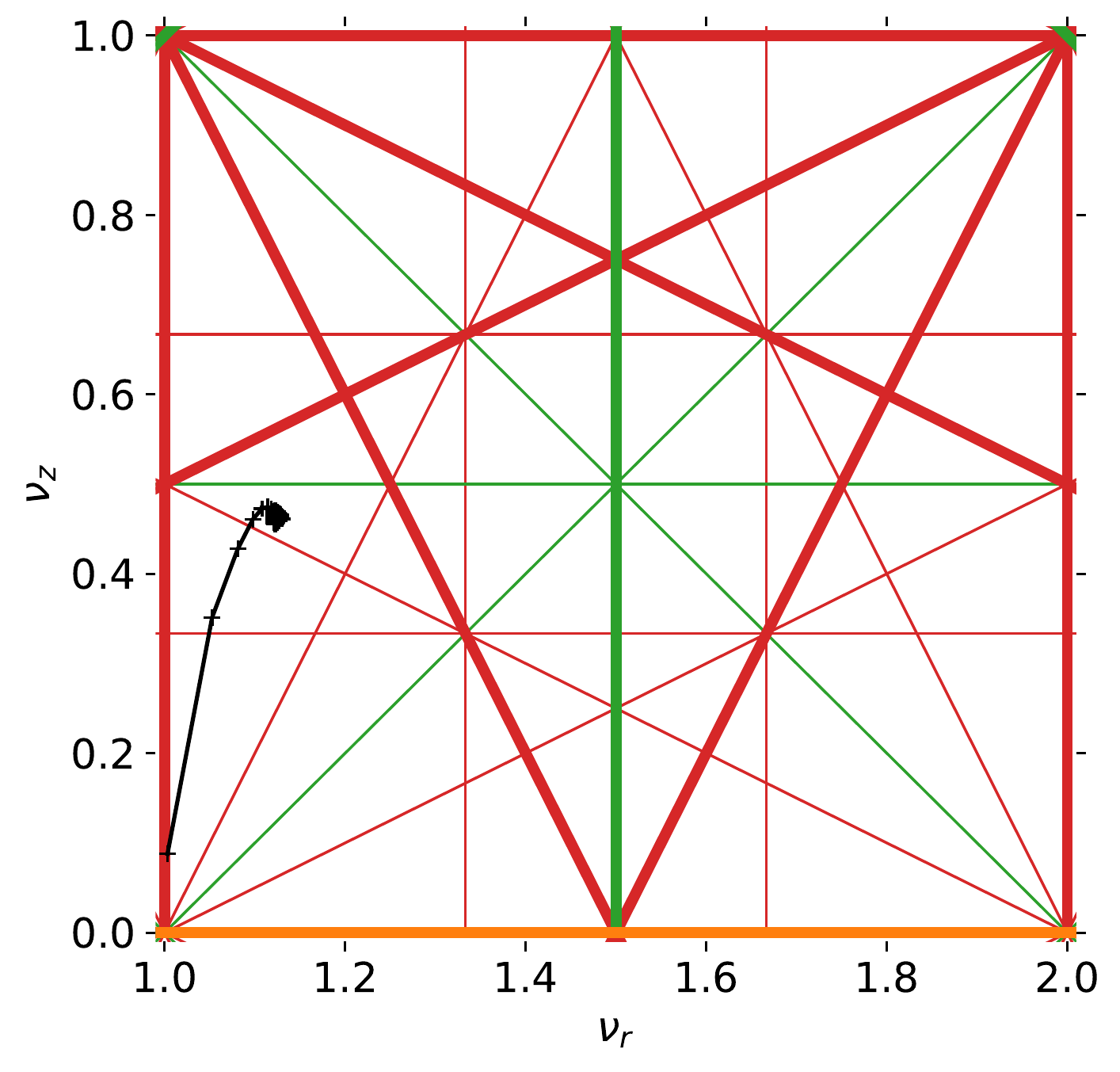}
    \centering
    \caption{Tunes obtained using the parameters in~\cref{tab:orb}. Betatron resonance conditions are shown up to 3$^\text{rd}$ order, with the structural resonances shown with thick lines and non-structural with thin lines.}
    \label{fig:tune-diag}
\end{figure}

To setup an optimization problem, one needs to choose an objective: we chose to minimize the RMS variation of the radial and vertical tunes over the acceleration range of the cyclotron. Now it is just a matter of letting some optimization routine (python \texttt{scipy.optimize.minimize} in our case) adjust the 8 degrees of freedom to best satisfy the objective function. Since the calculation of one set of tunes takes under a second, and the number of parameters to adjust in only 8, this process takes on the order of minutes. The result obtained in the case of a 3-sector cyclotron are presented in~\cref{tab:orb,fig:traj,fig:tunes,fig:tune-diag}. Note that, because of the constraint noted above, the tunes start at $\nu_r=1$ and $\nu_z=0$, but rapidly increase to $\nu_r\approx1.3$ and $\nu_z\approx0.47$ and remain there for the entire range of the machine (until $a=0.425\times\mathcal{R}_\infty$).
\begin{table}[h!b]
    \renewcommand{\arraystretch}{1.2}
    \begin{center}
        \begin{tabular}{ |c| l| l|}\hline
            $a/\mathcal{R}_\infty$ & $C$     & $\phi$/rad \\ \hline
            0.004                  & 0.      & 0.         \\ \hline
            0.14                   & 0.07297 & 0.5636     \\ \hline
            0.24                   & 0.08343 & 0.5687     \\ \hline
            0.35                   & 0.07129 & 0.4080     \\ \hline
            0.425                  & 0.04683 & 0.2383     \\ \hline
        \end{tabular}
        \caption{Orbit shape parameters for the compact cyclotron example.}
        \label{tab:orb}
    \end{center}
\end{table}
\FloatBarrier

% To go any further with the study, we need to choose the parameters that will turn this purely geometrical problem into an actual cyclotron design: these are the value of $\mathcal{R}_\infty$ and the mass and charge of the particles to accelerate. Because I am interested in the design of a cyclotron such as this~\cite{rao2020conceptual}), I choose $\mathcal{R}_\infty=5$\,m, and the particles to be H$_2^+$ ions. With this choice, one can now produce a field map of the cyclotron median plane from $r(a,\theta)$ using:
% \begin{equation}
%     B(r,\theta) =\frac{\beta(a(r,\theta))}{\sqrt{1-\beta^2(a(r,\theta))}}\frac{m}{q \rho(a(r,\theta),\theta)}\,,
% \end{equation}
% Note that the ``inverse'' function $a(r,\theta)$ must be constructed from $r(a,\theta)$ using some numerical method (I use a bicubic spline to construct this inverse function).

\begin{figure}[!htb]
    \centering
    \includegraphics*[width=0.8\columnwidth]{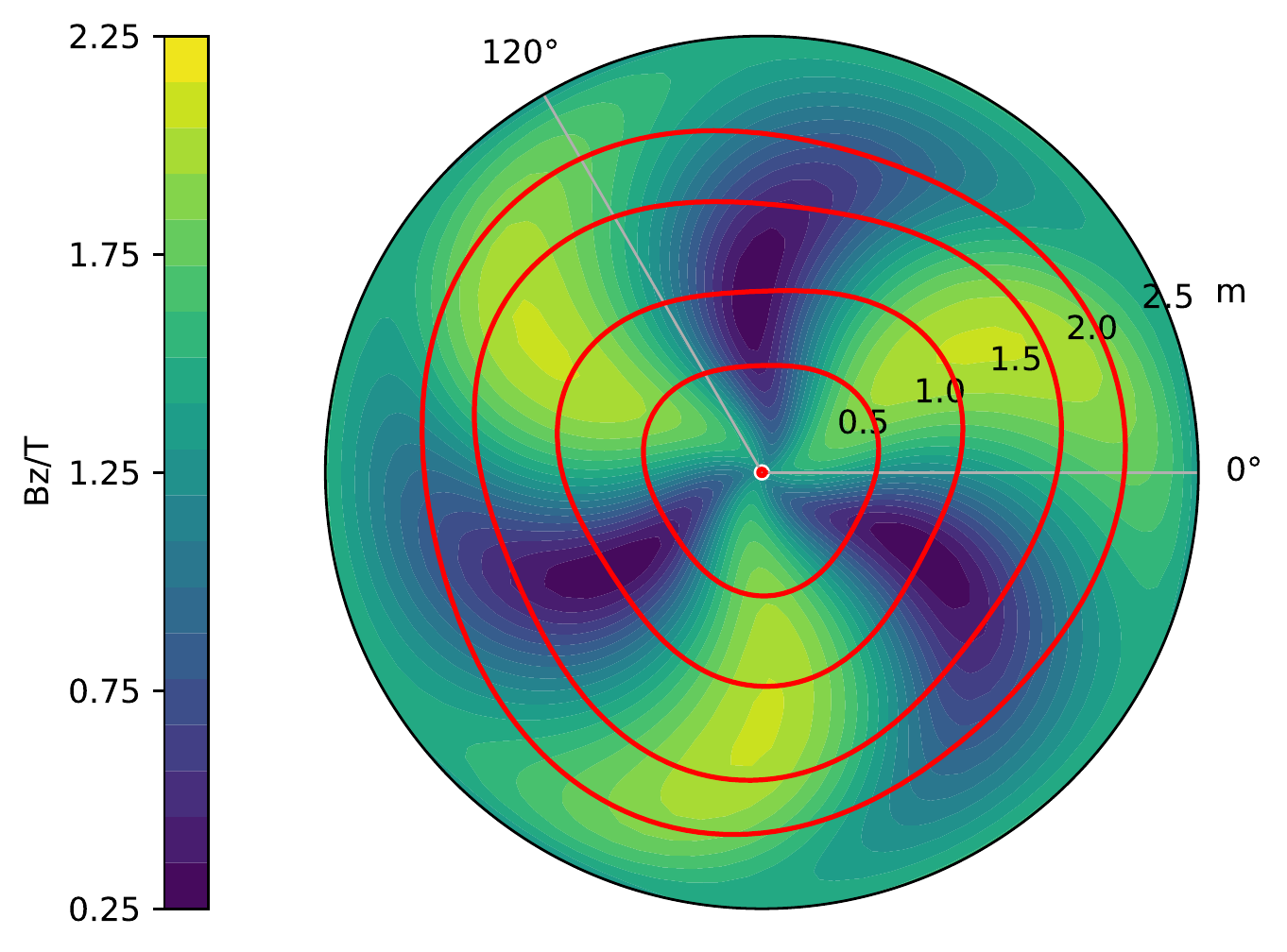}
    \centering
    \caption{Example of isochronous field map produced using \texttt{from\_orbit}. The 5 orbits shown in red were used to construct the spline function $r(a,\theta)$, which in turn was used to produce the magnetic field map.}
    \label{fig:traj}
\end{figure}

\begin{figure}[!htb]
    \centering
    \includegraphics*[width=0.7\columnwidth]{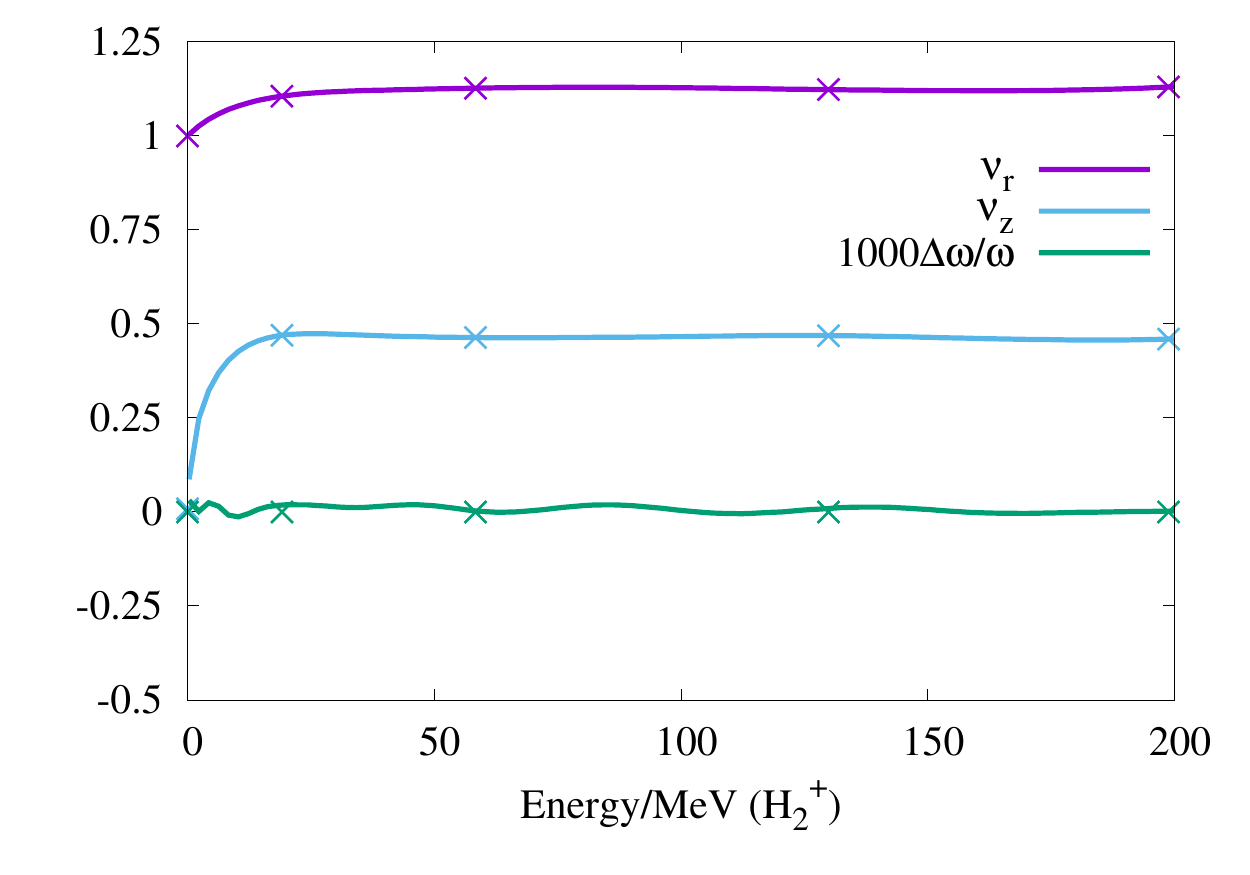}
    \centering
    \caption{Transverse tunes, and relative variation of the orbital frequency, plotted as a function of the H$_2^+$ beam energy for the example shown in~\cref{fig:traj}. Crosses show the results of calculation from \texttt{from\_orbit}. Solid lines are results obtained after extracting a magnetic field map from \texttt{from\_orbit}, and running it through the standard orbit code \texttt{CYCLOPS}\cite{gordon1984computation}.}
    \label{fig:tunes}
\end{figure}

\FloatBarrier
\subsection{Magnet Design: 200\,MeV H$_2^+$ Cyclotron }\label{sec:magnet}
\begin{figure}[!htb]
    \centering
    \includegraphics*[width=0.7\columnwidth]{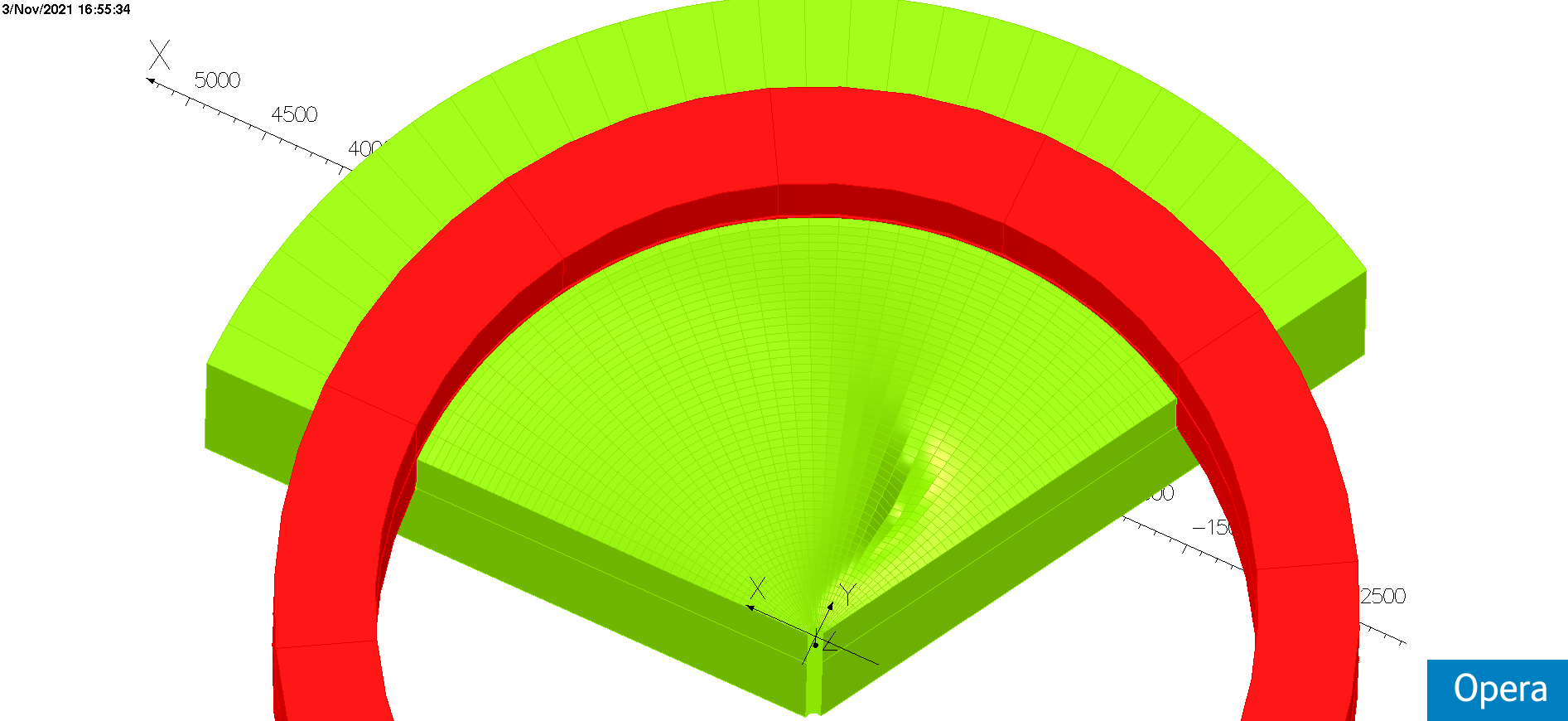}
    \centering
    \caption{View of the OPERA-3D magnet model.}
    \label{fig:opc}
\end{figure}

\begin{figure}[!htb]
    \centering
    \includegraphics*[width=0.8\columnwidth]{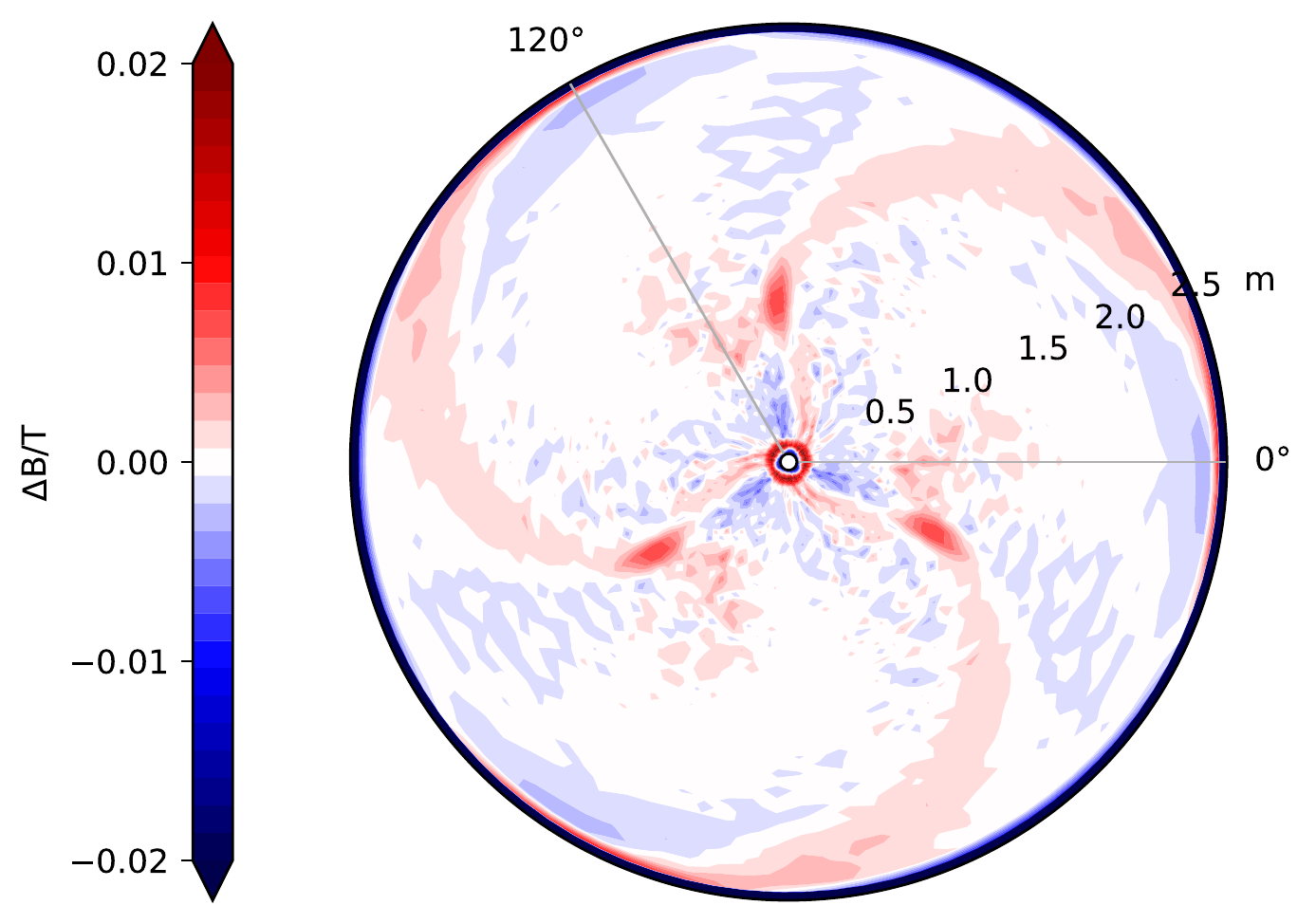}
    \centering
    \caption{Difference between the desired and the achieved magnetic field in the magnet mid-plane.}
    \label{fig:DBz}
\end{figure}

\begin{figure}[!htb]
    \centering
    \includegraphics*[width=0.7\columnwidth]{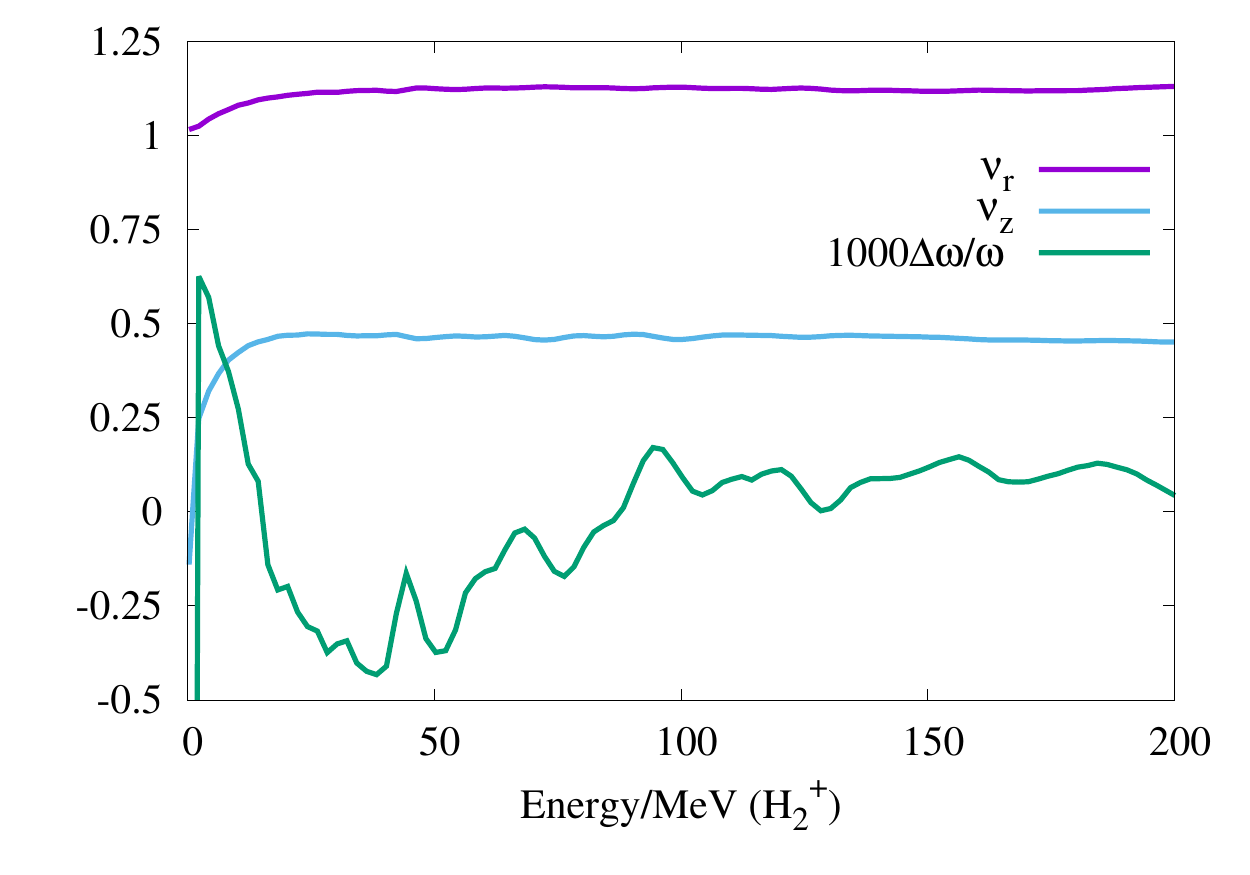}
    \centering
    \caption{Transverse tunes and  relative variation of the orbital frequency plotted as a function of the H$_2^+$ beam energy for the from a \texttt{CYCLOPS} simulation in the magnetic field map obtained from \texttt{OPERA-3d}.}
    \label{fig:tunes-OPERA}
\end{figure}

To design a magnet that produces the same field distribution as in~\cref{fig:traj}, we use a python script which produces \texttt{OPERA-3D} command input files, runs them, processes the solutions, and iterates. This entirely automatic procedure took about 70 iterations to converge to the result presented in~\cref{fig:opc,fig:DBz,fig:tunes-OPERA}, each iteration taking on the order of hours.

% \begin{figure}[!htb]
%     \centering
%     \includegraphics*[width=0.5\columnwidth]{tune-diag-OPERA}
%     \centering
%     \caption{Same tunes as in~\cref{fig:tunes-OPERA}, but plotted in the tune diagram, showing structural (thick lines) and non-structural (thin lines) resonance lines up to 3$^\text{rd}$ order.}
%     \label{fig:tune-diag-OPERA}
% \end{figure}

The tunes presented in~\cref{fig:tunes-OPERA} are in excellent agreement with~\cref{fig:tunes}. The relative variation of the orbital frequency plotted in~\cref{fig:tunes-OPERA} could still be improved, but is sufficient to allow isochronous acceleration to 200\,MeV with as little as 120\,kV of rf voltage per turn with a harmonic number of 1. This demonstrates that it is in principle possible to design an isochronous cyclotron with transverse tune constant over a wide momentum range.

One can however not build a practical cyclotron based on the magnet design presented in~\cref{fig:opc}: the magnet pole valleys are likely too shallow to install rf resonators. This means that we have to impose additional constraints on the shape of the orbits to obtain a more flat-top field distribution. It will lead us to use more harmonics in the Fourier series of~\cref{eq:rath1h}. This is what we explore in the next section.

\FloatBarrier
\section{High-Energy Ring Cyclotron Example}
Let's impose some constrains on $r(a,\theta)$ so that the orbits have long straight sections, like in a ring cyclotron with drift spaces between sectors. For this example we chose to use the same basic parameters -- number of sectors = 10, $\mathcal{R}_\infty$ = 28\,m, etc.\ -- of the proposed 0.8 to 2\,GeV CIAE ring cyclotron~\cite{zhang2019new,zhang2022this}.

\subsection{Smooth-Orbits With Straight Sections}
We cover the area between 0.8\,GeV and 2\,GeV using again 5 different orbits.
Each orbit is defined as a spline, with periodic boundary conditions. The splines are made to pass through a few points that are aligned, forcing each orbit to be straight over some minimum distance.  Each spline is parametrized with 4 constraints: the angular width and tilt of the straight section, plus the angular position and radius of the center of the reversed bend. We also introduce for each orbit an angular shift with respect to the innermost one. This sums up to a total of $5\times4 +4=24$ parameters.
We then calculate the Fourier transform of each of the 5 orbits as in~\cref{eq:fourier}, truncating at order $j=9$. Like was done in~\cref{seq:GeometricalDesign}, the values of each harmonics $C_j$ and $S_j$ are interpolated for intermediate values of $a$ using cubic splines.

We let the optimizer adjust these 24 free parameters, with the objective to minimize the RMS tune variation. We found a large number of solutions, and picked one with the tunes away from low-order betatron resonance lines, see~\cref{fig:tune-diag-2GeV}. The corresponding orbit shapes and field distribution are presented in~\cref{fig:field-2GeV,fig:field-orbits}.

\subsection{Constant-Tune 2\,GeV Cyclotron}

\begin{figure}
    \begin{center}
        \includegraphics[width=0.7\columnwidth]{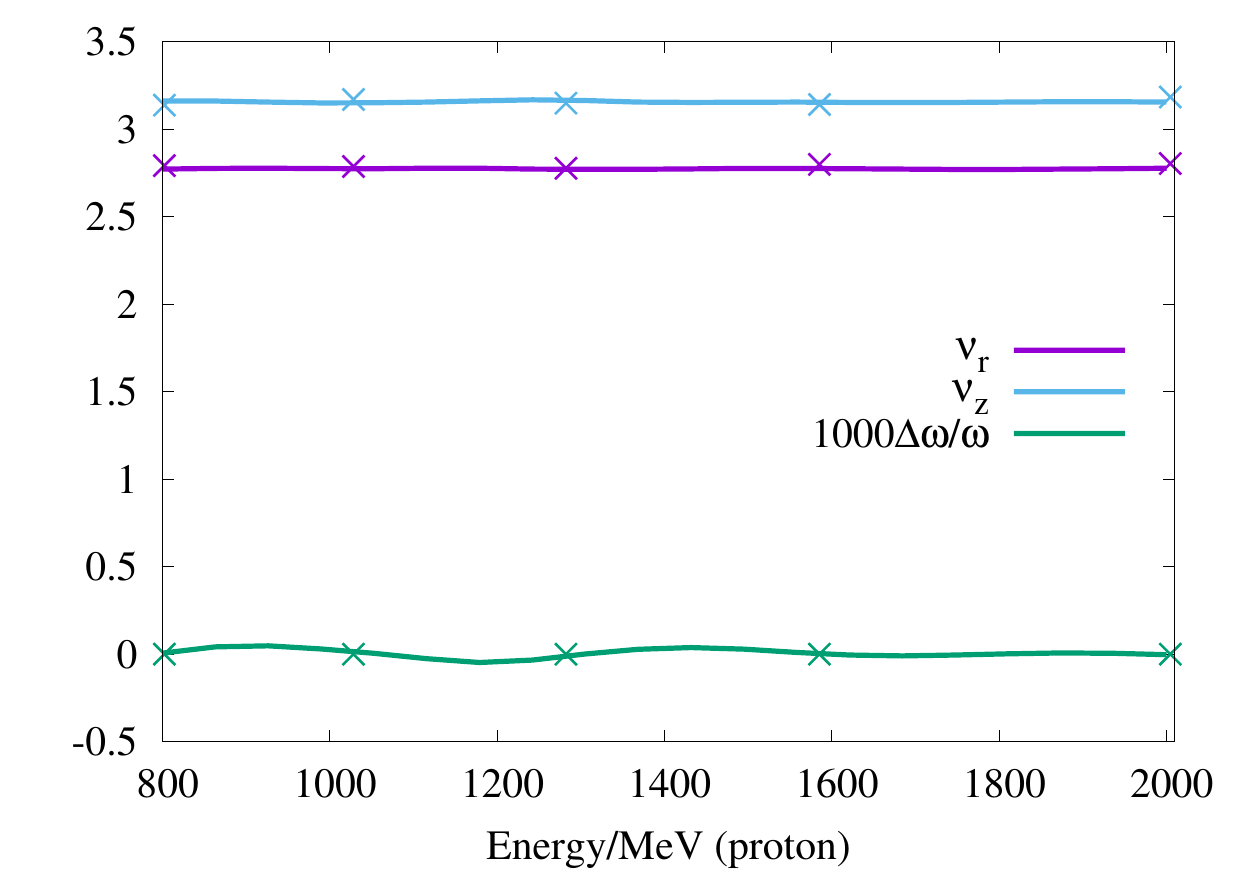}
        \caption{Transverse tunes, and relative variation of the orbital frequency, plotted as a function of the proton beam energy. Crosses show the results of calculation from \texttt{from\_orbit}. Solid lines are results obtained after extracting a magnetic field map from \texttt{from\_orbit}, and running it through the standard orbit code \texttt{CYCLOPS}.}\label{fig:tunes-2GeV}
    \end{center}
\end{figure}

\begin{figure}
    \begin{center}
        \includegraphics[width=0.7\columnwidth]{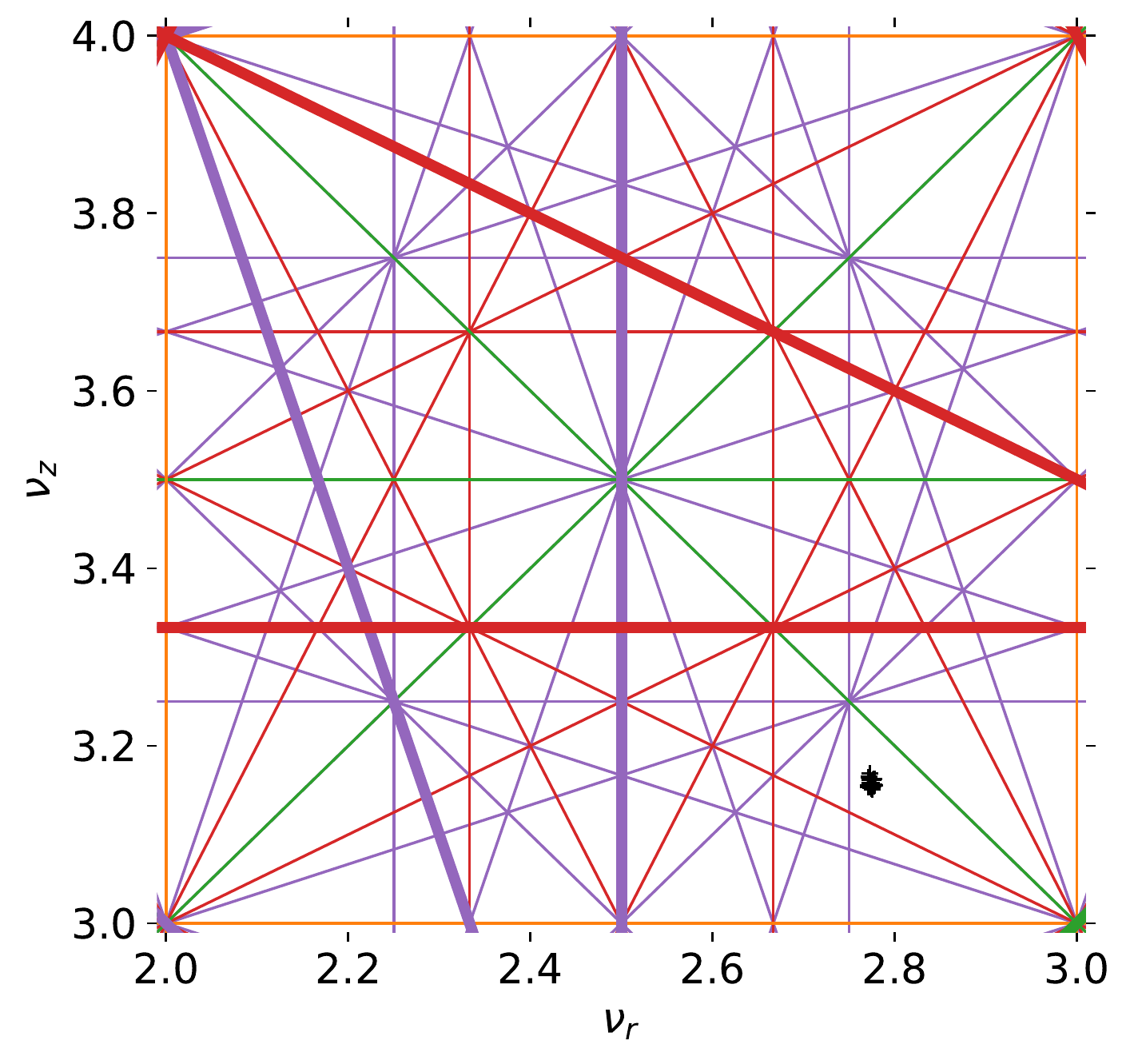}
        \caption{Tunes obtained using the field distribution shown in~\cref{fig:field-2GeV} are show in black, and appear as a small dot near the bottom right corner. Betatron resonance lines are shown up to 4$^\text{th}$ order, with the structural resonances shown with thick lines and non-structural with thin lines. }\label{fig:tune-diag-2GeV}
    \end{center}
\end{figure}

\begin{figure}
    \begin{center}
        \includegraphics[width=0.8\columnwidth]{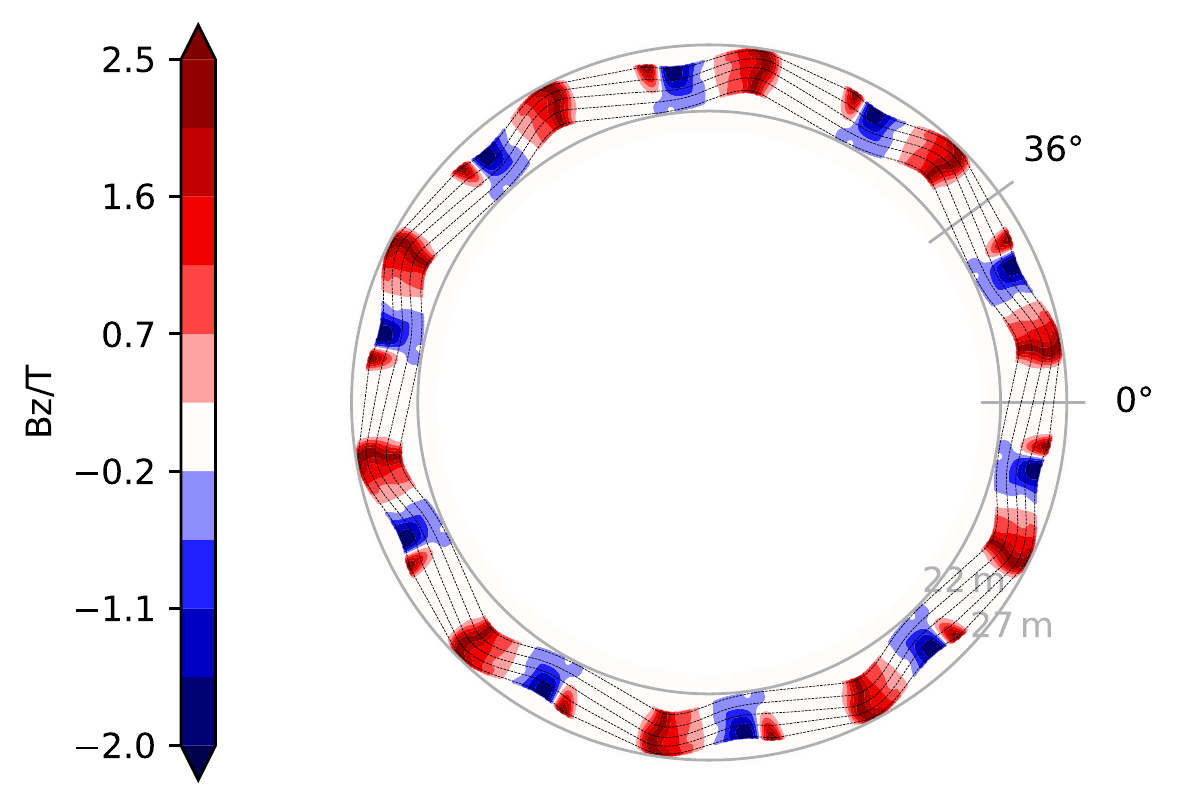}
        \caption{Magnetic field contours, with corresponding closed orbits, for a 10-sector 0.8 to 2\,GeV proton cyclotron. Five closed obits are shown with thin dotted lines. The radii of the inner orbit and outer orbit are given for scale. Note the regions of low magnetic field are available for installation of rf cavities, injection and extraction systems, etc.}\label{fig:field-2GeV}
    \end{center}
\end{figure}

\begin{figure}
    \begin{center}
        \includegraphics[width=0.8\columnwidth]{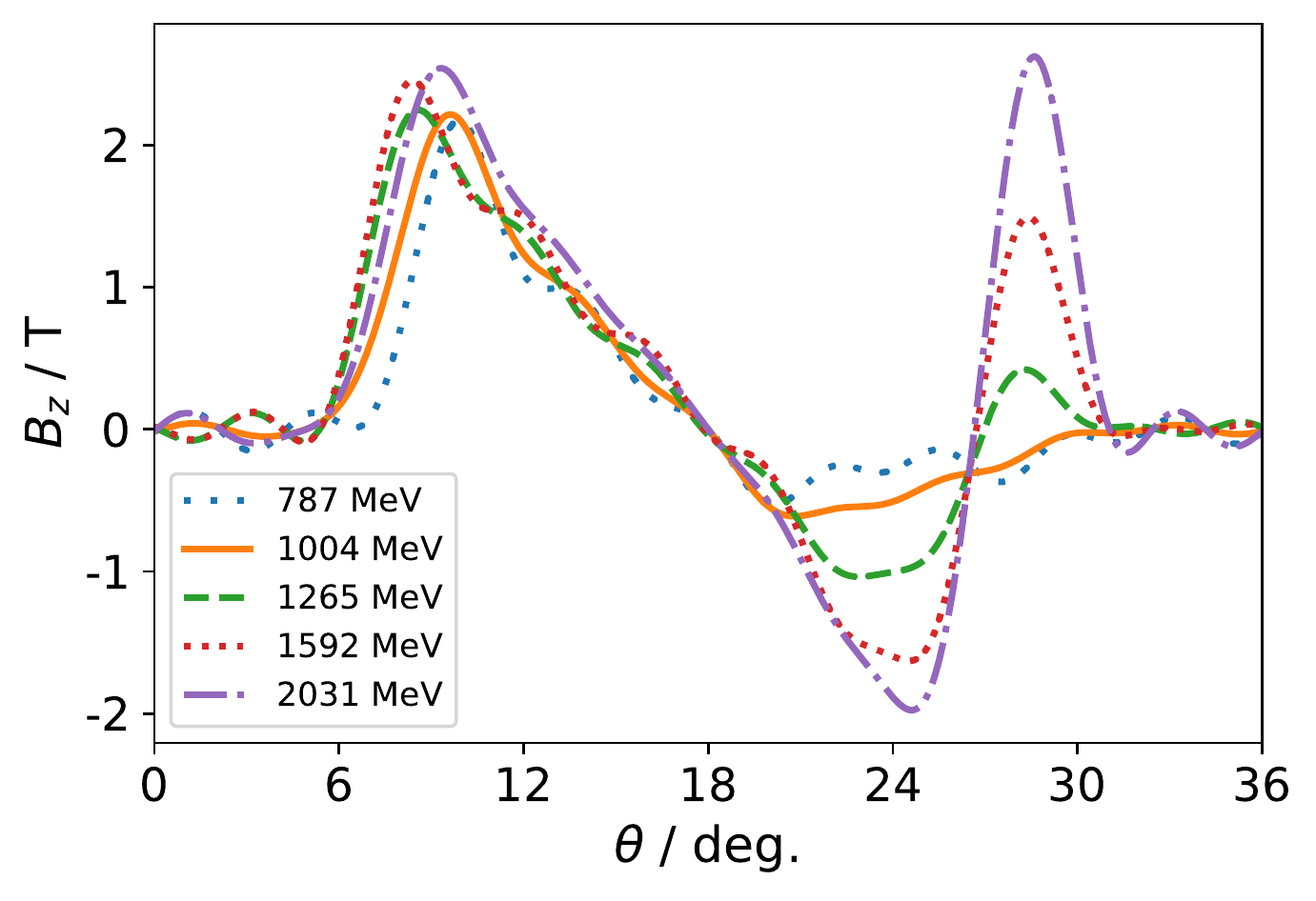}
        \caption{Magnetic field ($B_z$) along the 5 orbits shown in~\cref{fig:field-2GeV}. The proton energy for each orbit is shown in the legend.}\label{fig:field-orbits}
    \end{center}
\end{figure}

Both vertical and horizontal tunes are constant to better than 0.01, and the field is isochronous to a very high precision, see~\cref{fig:tunes-2GeV}. Particle tracking using the cyclotron code \texttt{CYCLOPS}\cite{gordon1984computation}, shown with solid lines on~\cref{fig:tunes-2GeV}, confirms this result. The corresponding magnetic field presents $\sim$9 degree wide low-field sections, see~\cref{fig:field-2GeV,fig:field-orbits}. The entire tune spread is barely visible on the tune diagram around the working point (2.78, 3.16), see~\cref{fig:tune-diag-2GeV}.

\FloatBarrier
\section{Conclusion}
We have demonstrated that cyclotron magnetic fields with constant tunes can be designed. Except for the central few orbits in a compact cyclotron, this avoids crossing betatron resonances. In ring cyclotrons, the tunes can be made completely constant, while being isochronous. Future work is required to convert the fields into actual steel and coil configurations.

The source code used to calculate tunes and generate fields maps for all the examples presented in this paper has been developed under a GPL-3 license and is available from:  \newline\url{https://gitlab.triumf.ca/tplanche/from-orbit}.

\FloatBarrier
% \bibliographystyle{unsrt}
% \bibliography{bib}

\begin{thebibliography}{10}

\bibitem{lawrence1934method}
EO~Lawrence.
\newblock Method and apparatus for the acceleration of ions, February~20 1934.
\newblock US Patent 1,948,384.

\bibitem{Craddock:1977xk}
MK~Craddock, EW~Blackmore, G~Dutto, CJ~Kost, GH~Mackenzie, and P~Schmor.
\newblock {Improvements to the Beam Properties of the TRIUMF Cyclotron}.
\newblock {\em IEEE Trans.\ Nucl.\ Sci.\ (Proc.\ PAC'77)}, 24:1615--1617, 1977.

\bibitem{gordon1968orbit}
MM~Gordon.
\newblock Orbit properties of the isochronous cyclotron ring with radial
  sectors.
\newblock {\em Annals of Physics}, 50(3):571--597, 1968.

\bibitem{TRI-BN-82-12}
Werner Joho.
\newblock Simulating cyclotron sector magnets with the computer program
  {TRANSPORT}.
\newblock Technical Report TRI-BN-82-12, {TRIUMF}, 1982.

\bibitem{TRI-BN-82-20}
J.~Botman.
\newblock Ring cyclotrons with return field gullies.
\newblock Technical Report TRI-BN-82-20, {TRIUMF}, 1982.

\bibitem{craddock2008cyclotrons}
MK~Craddock and KR~Symon.
\newblock Cyclotrons and fixed-field alternating-gradient accelerators.
\newblock {\em Reviews of accelerator science and technology}, 1(01):65--97,
  2008.

\bibitem{CourantSnyder1958}
E.D Courant and H.S Snyder.
\newblock Theory of the alternating-gradient synchrotron.
\newblock {\em Annals of Physics}, 3(1):1 -- 48, 1958.

\bibitem{TRI-DN-05-06}
R.~Baartman.
\newblock {Linearized Equations of Motion in Magnet with Median Plane
  Symmetry}.
\newblock Technical Report TRI-DN-05-06, TRIUMF, 2005.

\bibitem{sacherer1971rms}
Frank~J Sacherer.
\newblock Rms envelope equations with space charge.
\newblock {\em IEEE Transactions on Nuclear Science}, 18(3):1105--1107, 1971.

\bibitem{TRI-DN-70-54-Add4}
Colleen Meade.
\newblock Changes to the main magnet code.
\newblock Technical Report TRI-DN-70-54-Addendum4, {TRIUMF}, 1971.

\bibitem{planche2019designing}
Thomas Planche.
\newblock {Designing Cyclotrons and Fixed Field Accelerators From Their
  Orbits}.
\newblock In {\em {Proc.\ of Int.\ Conf.\ on Cyclotrons and their Applications
  (Cyclotrons'19)}}, page FRB01, 2019.

\bibitem{zhang2022asurement}
L.~Zhang, Y.-N. Rao, R.~Baartman, Y.~Bylinski, and T.~Planche.
\newblock Redundant field survey data of cyclotron with imperfect median plane.
\newblock {\em in this issue}, 2022.

\bibitem{gordon1984computation}
MM~Gordon.
\newblock Computation of closed orbits and basic focusing properties for
  sector-focused cyclotrons and the design of {CYCLOPS}.
\newblock {\em Part.\ Accel.}, 16:39--62, 1984.

\bibitem{planche2022intensity}
T.~Planche, R.~Baartman, Y.-N. Rao, and L.~Zhang.
\newblock Intensity limit in compact h$^-$ and h$_2^+$ cyclotrons.
\newblock {\em in this issue}, 2022.

\bibitem{zhang2019new}
TJ~Zhang, Ming Li, Tianjian Bian, Chuan Wang, Zhiguo Yin, Shilun Pei, and
  Shizhong An.
\newblock A new solution for cost effective high average power {(2\,GeV 6\,MW)}
  proton accelerator and its {R\&D} activities.
\newblock In {\em Proc.\ Cyclotrons}, pages 334--339, 2019.

\bibitem{zhang2022this}
Tianjue Zhang et~al.
\newblock Innovative research and development activities on high energy and
  high current isochronous proton accelerator.
\newblock {\em in this issue}, 2022.

\end{thebibliography}

\end{document}